\documentclass[12pt,a4paper]{article}
\pdfoutput=1

\usepackage[utf8]{inputenc}
\usepackage[T1]{fontenc}
\usepackage[margin=2.0cm]{geometry}
\usepackage[pdfstartview={FitH},bookmarks=false,linktoc=page,colorlinks=true,linkcolor=blue,citecolor=blue,urlcolor=blue]{hyperref}
\usepackage[numbered]{bookmark}
\usepackage{mathtools}
\usepackage{amssymb}
\usepackage{graphicx}
\usepackage[font=small,labelfont=bf]{caption}
\usepackage{authblk}
\usepackage{cite}
\usepackage{dsfont}
\usepackage{xcolor}
\usepackage[titles]{tocloft}
\usepackage{float}
\usepackage{subcaption}
\usepackage{color}
\usepackage{tikz}
\usepackage{ifthen}
\usepackage[warn]{textcomp}%for the number sign
\numberwithin{equation}{section}
\allowdisplaybreaks
\usepackage{multirow}

\begin{document}

%%%%%%%%%%%%%%%%%%%%%%%%%%%%%%%%%%%%%%%%%%%%
\title{\vspace{2cm}\textbf{Thermodynamic Information Geometry and Complexity Growth of Warped AdS Black Hole and the WAdS$_3$/CFT$_2$ Correspondence}\vspace{1cm}}
%%%%%%%%%%%%%%%%%%%%%%%%%%%%%%%%%%%%%%%%%%%%

\author[a,b]{H. Dimov}
\author[a,c]{R. C. Rashkov}
\author[a,b]{T. Vetsov}

\affil[a]{\textit{Department of Physics, Sofia University,}\authorcr\textit{5 J. Bourchier Blvd., 1164 Sofia, Bulgaria}

\vspace{-10pt}\texttt{}\vspace{0.0cm}}

\affil[b]{\textit{The Bogoliubov Laboratory of Theoretical Physics, JINR,}\authorcr\textit{141980 Dubna,
Moscow region, Russia}

\vspace{-10pt}\texttt{}\vspace{0.0cm}}

\affil[c]{\textit{
Institute for Theoretical Physics, Vienna University of Technology,}\authorcr\textit{Wiedner Hauptstr. 8–10, 1040 Vienna, Austria}

\vspace{10pt}\texttt{h\_dimov,rash,vetsov@phys.uni-sofia.bg}\vspace{0.1cm}}
\date{}
\maketitle

\begin{abstract}
We study the thermodynamic properties of warped AdS$_3$ black hole within the framework of thermodynamic information geometry. Our analysis focuses on finding the set of proper thermodynamic Riemannian metrics on the space of equilibrium states, together with the conditions for local and global thermodynamic stability. We use our findings to constrain the values of left and right central charges from the dual CFT theory and the parameters of the bulk gravitational theory.

\end{abstract}

\vspace{0.5cm}
\vspace{0.5cm}
\thispagestyle{empty}

\noindent\rule{\linewidth}{0.75pt}
\vspace{-0.8cm}\tableofcontents
\noindent\rule{\linewidth}{0.75pt}

\section{Introduction}\label{sec: Introduction}
In the past few decades the AdS/CFT correspondence  revealed important perturbative and non-perturbative phenomena in various classical and quantum models. One of its important feature is that it relates a classical gravitational theory in the bulk of space-time to a quantum theory without gravity on a lower-dimensional boundary, and vice versa. On the other hand, the correspondence is also a duality between weak/strong coupling regimes of both theories. 
\\
\indent A specific example of such duality is given by the correspondence between a warped three-dimensional anti-de Sitter black hole and a two-dimensional conformal field theory. In this case, the warped AdS black hole is a stable vacuum solution of the 3$d$ topological massive gravity (TMG), described by the action \cite{Anninos:2008fx}:
\begin{equation}\label{eq_TMG_action}
{I_{TMG}} = \frac{1}{{16\,\pi \,G}}\,\int_{\cal M} {{d^3}} x\,\sqrt { - g} \,\left( {R + \frac{2}{{{L ^2}}}} \right) + \frac{1}{\mu }\,{I_{CS}}+\int_{\partial \mathcal{M}}B\,.
\end{equation}
Here, $I_{CS}$ is the gravitational Chern-Simons action,
\begin{equation}
{I_{CS}} = \frac{1}{{32\,\pi \,G}}\,\int_{\cal M} {{d^3}} x\,\sqrt { - g} \,{\varepsilon ^{\lambda \mu \nu }}\,\Gamma _{\lambda \sigma }^r\left( {{\partial _\mu }\Gamma _{r\nu }^\sigma  + \frac{2}{3}\,\Gamma _{\mu \tau }^\sigma \,\Gamma _{\nu r}^\tau } \right)\,,
\end{equation}
and the coupling $\mu$ is the mass of the graviton, $\varepsilon^{\lambda \mu\nu}=\epsilon^{\lambda \mu\nu}/\sqrt{-g}$, $\epsilon^{012}=+1$. The boundary term,
\begin{equation}
B = \frac{1}{{32\,\pi \,G}}\,{\epsilon _{ABC}}\,{\omega ^{AB}}\wedge{e^C}\,,
\end{equation}
was introduced in \cite{Miskovic:2009kr} to make the variational
principle well-defined. For every value of the coupling $\mu$ TMG has a classical
AdS$_3$ solution with a radius $L$. The  only stable case is defined by the condition $\mu\,L = 1$, which leads to a non-negative energy of the gravitons. In this case, it is also possible to construct a consistent
quantum theory of the so-called chiral gravity \cite{Li:2008dq}. 
\\
\indent However, if we consider non-chiral values of $\mu\,L$, we can construct other stable TMG vacua, namely warped backgrounds. They are discrete quotients by elements of $SL(2, R) \times U (1)$ of warped AdS$_3$ space. In this particular case, the group elements of the quotient select the left and the right temperatures of the corresponding boundary CFT. With a suitable choice of the central charges the density of states in the boundary CFT exactly matches the Bekenstein-Hawking entropy of the corresponding black hole, thus a duality between both theories can be conjectured. It is also worth mentioning that  warped AdS$_3$ solutions arise in a number of contexts besides TMG, see e.g. \cite{Gurses:1994bjn, Rooman:1998xf, Duff:1998cr, Israel:2003ry, Andrade:2005ur, Bengtsson:2005zj, Banados:2005da, Son:2008ye, Balasubramanian:2008dm}.
\\
\indent Recently, the holographic principle was used to describe the complexity of a quantum state via bulk theory computations. In this case, there are two conjectures. The first one is called ''Complexity equals Action'' (CA) \cite{Brown:2015lvg, Brown:2015bva}, which states that in order to compute the quantum computational complexity of a holographic state, one can calculate the on-shell action on the so called “Wheeler-De Witt” patch:
\begin{equation}
{C_A}(\Sigma ) = \frac{{{I_{WDW}}}}{{\pi \hbar }},
\end{equation}
where $\Sigma$ is the time slice intersection of the asymptotic boundary and the Cauchy surface in the bulk. The second proposal, known as ''Complexity equals Volume'' (CV) \cite{Susskind:2014rva, Susskind:2014moa, Stanford:2014jda}, relates the complexity of the boundary states with the volume $V$ of a maximal slice behind the event horizon, i.e.
\begin{equation}
{C_V}(\Sigma ) = \mathop {\max }\limits_{\Sigma  = \partial B} \left[ {\frac{{V(B)}}{{{G_N}\ell}}} \right]\,.
\end{equation}
A nice feature of the VC conjecture is that the maximal volume  naturally grows at a rate proportional to
the product between the temperature $T$ and the entropy $S$ of the black hole.
Moreover, the idea that the growth of complexity can be interpret as a computation naturally invokes Lloyd's bound \cite{Lloyd2000} on the rate of computation for systems with energy $M$:
\begin{equation}
\frac{{dC}}{{dt}} \le \frac{{2\,M}}{{\pi \,\hbar }}\,.
\end{equation}
\indent Furthermore, several other information-theoretic concepts have been fruitfully applied to the investigation of
fundamental properties of various gravitational systems, namely entanglement entropy \cite{Ryu:2006bv, Casini:2011kv, Lewkowycz:2013nqa, Dimov:2016vvl, Dimov:2017ryz}, Fisher information metric \cite{Dimov:2016vvl, Dimov:2017ryz}, and thermodynamic information geometry (TIG) \cite{Aman:2003ug, Shen:2005nu, Cai:1998ep, Aman:2007pp, Sarkar:2006tg, Mansoori:2016jer, Mansoori:2014oia, Mansoori:2013pna, Ruppeiner:2018pgn, Vetsov:2018dte}. The latter can be considered as a specific thermodynamic limit of the quantum Fisher metric. The common belief is that the classical space-time geometry has the capacity to
encode important properties of the dual quantum system. Due to the lack of consistent theory of quantum gravity a case by case study is required. Here, we will focus on the thermal properties of the warped $AdS_3$ black hole via the methods of TIG. 
\\
\indent Thermodynamic information geometry was first introduced by F. Weinhold \cite{doi:10.1063/1.431689} and later by G. Ruppeiner \cite{RevModPhys.67.605}. Weinhold showed that the empirical laws of equilibrium thermodynamics can be brought into correspondence with the mathematical axioms of an abstract metric space. In his approach, Weinhold used the Hessian of the internal energy with respect to the extensive parameters of the system in order to introduce a Riemannian metric on the space of macro states,
\begin{equation}
g_{ab}^{(W)} = {\partial _a}{\partial _b}U(E^c )\,.
\end{equation}
Here $E^c$ are the other extensive parameters of the system besides $U$. On the other hand, Ruppeiner developed his geometric approach within fluctuation theory, where one implements the entropy $S(E^c)$ as a thermodynamic potential in order to define a Hessian metric structure on the state space statistical manifold:
\begin{equation}
{g^{(R)}_{ab}} =  - {\partial _a}{\partial _b}{S}(E^c)\,.
\end{equation}
Here $E^c$ are the other extensive parameters of the system besides $S$.
\\
\indent The importance of using Hessian metrics on the equilibrium manifold is best understood, when one considers small fluctuations of the thermodynamic
potential. The potential is extremal at each equilibrium point, while the second moment
of the fluctuation turns out to be directly related to the components of the corresponding Hessian metric. As it turned out, both metric approaches are conformally related to each other with the temperature being the conformal factor,
\begin{equation}
ds^2_{(R)}=\frac{1}{T}\,ds^2_{(W)}\,.
\end{equation}
\indent Although Weinhold and Ruppeiner metrics have been successfully applied to describe the phase structure of condensed matter systems, when utilized for black holes they do not often agree with each other. One of the reasons is due to the fact that Hessian metrics are not Legendre invariant, thus they do not preserve the geometric structure when a different thermodynamic potential is used for the description of the equilibrium states. In order to make things Legendre invariant, H. Quevedo considered the $(2n + 1)$-dimensional thermodynamic phase space, spanned by the thermodynamic potential $\Phi$, the set of extensive
variables $E^a$, and the set of intensive variables $I^a $, $a = 1,\dots, n$. In Ref.  \cite{Quevedo:2017tgz} it was found that the general metric on
the equilibrium state space can be written in the form
\begin{equation}
{g^{I,\,II}} = {\Omega _\Phi }\,\Phi ({E^c})\,\chi _a^{\,\,b}\frac{{{\partial ^2}\Phi }}{{\partial {E^b}\,\partial {E^c}}}\,d{E^a}\,d{E^c}\,,
\end{equation}
where $\chi^{\,\,b}_a=\chi_{af}\,\delta^{fb}$ is a constant diagonal matrix and $\Omega_\Phi\in\mathbb{R}$ is the degree of generalized homogeneity, $\Phi ({\lambda ^{{\beta _1}}}\,{E^1}, \ldots ,{\lambda ^{{\beta _N}}}\,{E^N})=$ ${\lambda ^{{\Omega _\Phi }}}\,\Phi ({E^1}, \ldots ,{E^N}),\,{\beta _a} \in \mathbb{R}$. In this case, the Euler identity for homogeneous functions is given by
\begin{equation}\label{eqEulerIdentity}
{\beta _{ab}}\,{E^a}\,\frac{{\partial \Phi }}{{\partial {E^b}}} = {\Omega _\Phi }\,\Phi \,,
\end{equation}
where ${\beta _{ab}} = {\mathop{\rm diag}\nolimits} ({\beta _1},\,{\beta _2}, \ldots ,{\beta _N})$. From the first law $ d\Phi = I_a\,dE^a$, one notes that $I_a =\partial \Phi/\partial E^a$. When $\beta_{ab}=\delta_{ab}$, one returns to the standard Euler's identity. If we choose $\beta_{ab}=\delta_{ab}$, for complicated systems this may lead to some non-trivial conformal factor, which is no longer proportional to the potential $\Phi$. On the other hand, if we set $\chi_{ab}=\delta_{ab}$, the resulting
metric $g^I$ can be used to investigate systems with at least one first-order phase transition. Alternatively, the choice $\chi_{ab}=\eta_{ab}={\rm{diag}}(-1,1,\dots,1)$, leads to a different metric $g^{II}$, which applies to systems with second-order phase transitions.
\\
\indent Consequently, one can show that the corresponding scalar thermodynamic curvature encodes information about the phase structure of the system. As suggested by G. Ruppeiner in Ref. \cite{RevModPhys.67.605}, the Ricci information curvature $R_I$ is related to the correlation volume of the system. This association follows from the idea that it will be less probable to fluctuate from one equilibrium thermodynamic state to the other, if the distance between the points on the statistical manifold, which correspond to these states, is larger. Moreover, the sign of the scalar curvature can be linked to the nature of the inter-particle interactions, \cite{2010AmJPh..78.1170R}. For example, if $R_I=0$, the interactions are
absent and the system is free. For positive curvature the interactions are repulsive, while for negative curvature the interactions are attractive. 
\\
\indent Finally, the scalar curvature on the equilibrium manifold can also be used to measure the stability of the physical system under thermodynamic perturbations. In particular, the information curvature approaches infinity in the vicinity of critical points, where phase transitions occur \cite{Janyszek:1989zz}. Moreover, the curvature of the information metric tends to diverge not only at the critical points of phase transitions, but on entire regions of points on the statistical manifold, called spinodal curves. The latter can be used to discern physical from non-physical situations. 
\\
\indent In this paper we investigate the properties of warped AdS$_3$ black hole solution of TMG within the framework of thermodynamic information geometry. The text is structured as follows. In section \ref{sec2} we introduce the warped AdS$_3$ black hole and the relevant thermodynamic quantities. In Section \ref{sec3} we calculate the heat capacity of the WAdS$_3$ black hole in different parameter spaces and analyze 
the conditions for local thermodynamic stability of the system. In Section \ref{sec4} we calculate the Hessian thermodynamic metrics on the space of equilibrium states and show that they fail as viable Riemannian metrics. Also, in this section we identify the regions of global thermodynamic stability in canonical and grand canonical ensemble. In Section \ref{sec5} we investigate the properties of two conformally related Legendre invariant metrics, namely Quevedo and HPEM information metrics. We show that they successfully reproduce the Davies phase transition points and thus can be used to describe the space of equilibrium states. In Section \ref{sec6} we investigate a third approach for introducing thermodynamic metrics on the equilibrium manifold by defining various conjugate thermodynamic potentials. By construction, the thermodynamic metrics in this approach are designed to match the Davies transition points of the heat capacity, without any additional singularities. We confirm that this is also valid for the WAdS$_3$ case. In Section \ref{sec7} we calculate the complexity growth of the warped solution and show that it is compatible with the conditions for thermodynamic stability. Furthermore, we find that the Lloyd's bound on the complexity growth leads to a lower bound on the mass of the black hole. Finally, in Section \ref{sec8} we make brief comments on our results.

\section{Warped AdS$_3$ black hole and its thermodynamics}\label{sec2}

The warped AdS$_3$ solution of (\ref{eq_TMG_action}) is given by the following metric \cite{Anninos:2008fx}
\begin{align}\label{eq_WAdS3_Metric}
\frac{{d{s^2}}}{{{L ^2}}} &= d{t^2} + \frac{{d{r^2}}}{{({\nu ^2} + 3)\,(r - {r_ + })\,(r - {r_ - })}} + \left( {2\,\nu \,r - \sqrt {{r_ + }\,{r_ - }\,({\nu ^2} + 3)} } \right)\,dt\,d\theta \\\nonumber
 &+ \frac{r}{4}\,\left( {3\,({\nu ^2} - 1)\,r + ({\nu ^2} + 3)\,({r_ + } + {r_ - }) - 4\,\nu \,\sqrt {{r_ + }\,{r_ - }\,({\nu ^2} + 3)} } \right)\,d{\theta ^2}\,,
\end{align}
where ${r}\in[0,\infty]$, $t\in[-\infty,\infty]$ and $\theta \sim \theta + 2\pi$. The horizons are located at ${r}_{+}$ and ${r}_{-}$, where $1/g_{{r}{r}}$ as well as the determinant of the $(t,\theta)$ metric vanishes. Here, we also introduced the parameter $\nu  = \mu \,L /3$. Notice that (\ref{eq_WAdS3_Metric}) reduces to the BTZ black hole in a rotating frame, when
$\nu^2 = 1$. For $\nu^2 > 1$, we have physical black holes, as long
as $r_{+}$ and $r_{-}$ stay positive. For $\nu^2 < 1$, we always encounter closed timelike curves and such geometries will not be considered. The Ricci curvature and the Kretschmann scalar invariant of the metric are
\begin{equation}\label{eq_Physical_Invariants}
R = -\frac{6}{{{L^2}}}\,,\qquad K = 6 \,\frac{{3 - 2\,{\nu ^2} + {\nu ^4}}}{{{L^4}}}.
\end{equation}
Nonphysical situations occur at $L=0$ and $\nu\to \pm \infty$, thus it is natural to consider $L>0$ and finite $\nu^2>1$ for the non-chiral case.
\\
\indent The entropy $S$ of the warped AdS black hole and the ADT conserved charges $M$ and $J$ are given by
\begin{equation}
S = \frac{{\pi \,L}}{{24\,\nu \,G}}\,\left( {(9\,{\nu ^2} + 3)\,{r_ + } - ({\nu ^2} + 3)\,{r_ - } - 4\,\nu \,\sqrt {({\nu ^2} + 3)\,{r_ + }\,{r_ - }} } \right)\,,
\end{equation}
\begin{equation}
M = \frac{{({\nu ^2} + 3)}}{{24\,G}}\,\left( {{r_ + } + {r_ - } - \frac{1}{\nu }\,\sqrt {{r_ + }\,{r_ - }\,({\nu ^2} + 3)} } \right)\,,
\end{equation}
and
\begin{equation}
J = \frac{{\nu \,L\,({\nu ^2} + 3)}}{{96\,G}}\,\left[ {{{\left( {{r_ + } + {r_ - } - \frac{1}{\nu }\,\sqrt {{r_ + }\,{r_ - }\,({\nu ^2} + 3)} } \right)}^2}} \right.\left. { - \frac{{(5\,{\nu ^2} + 3)}}{{4\,{\nu ^2}}}\,{{({r_ + } - {r_ - })}^2}} \right]\,.
\end{equation}
One can also compute the Hawking temperature and the angular velocity:
\begin{equation}\label{eqTandOmegainRSpace}
T = \frac{{\left( {{\nu ^2} + 3} \right)\,({{{r}}_ + } - {{{r}}_ - })}}{{4\,\pi \,L\left( {2\,\nu \,{{{r}}_ + } - \sqrt {\left( {{\nu ^2} + 3} \right){{{r}}_ + }\,{{{r}}_ - }} } \right)}}\,,\quad \;\;\:\Omega  = \frac{2}{{L\,\left( {2\,\nu \,{{{r}}_ + } - \sqrt {\left( {{\nu ^2} + 3} \right){{{r}}_ + }\,{{{r}}_ - }} } \right)}}\,.
\end{equation}
In this case, the first law of thermodynamics holds,
\begin{equation}\label{eq_The_First_Law_of_TD}
dM = T\,dS + \Omega \,dJ\,.
\end{equation}
Here, it is relevant to comment on whether or not the thermodynamic quantities can become zero. Considering $r_{+}\geq r_{-}$, one can reach $S=0$ only if $-1\leq \nu \leq 1$, which is excluded. The same condition holds for $M=0$. Thus, we consider only  positive $S>0,\,M>0$. From Eq. (\ref{eqTandOmegainRSpace}) one notes that the angular velocity $\Omega$ cannot be zero, while the temperature $T$ is zero for coincident horizons, $r_{+}=r_{-}$. On the other hand, the laws of thermodynamics forbid us from ever reaching the absolute zero, thus $T>0$. The only quantity, which can become zero within $\nu^2>1$, is $J$. Without loss of generality, we will consider only non-negative values of the angular charge, $J\geq 0$.
\\
\indent Instead of $r_{+}$ and $r_{-}$, we can work with the left and right temperatures, namely
\begin{equation}
{T_R} = \frac{{({\nu ^2} + 3)\,({r_ + } - {r_ - })}}{{8\,\pi \,L}}\,,\qquad {T_L} = \frac{{({\nu ^2} + 3)}}{{8\,\pi \,L}}\,\left( {{r_ + } + {r_ - } - \frac{{\sqrt {({\nu ^2} + 3)\,{r_ + }\,{r_ - }} }}{\nu }} \right)\,,
\end{equation}
and the left and right central charges, 
\begin{equation}\label{eq_CFT_Left_and_Right_Central_Charges}
{c_R} = \frac{{(5\,{\nu ^2} + 3)\,L}}{{G\,\nu \,({\nu ^2} + 3)}}\,,\qquad {c_L} = \frac{{4\,\nu \,L}}{{G\,({\nu ^2} + 3)}}\,,\qquad {c_L} - {c_R} =  - \frac{L}{{G\,\nu }}\,,
\end{equation}
of the dual CFT$_2$ theory. Although we are going to consider only positive central charges throughout the paper, which lead to unitary CFTs, one should keep in mind that negative charges can play vital role in anomaly cancellations, when considering the total central charge. From Eqs. (\ref{eq_CFT_Left_and_Right_Central_Charges}), under the requirement of positive central charges and $\nu^2>1$, one can restrict only to $\nu>1$. Therefore, it immediately follows that $c_L<L$ and $c_R<2 \,L$. For large $\nu\to\infty$ one has vanishing central charges, which is physically excluded due to the divergence of the Kretschmann  invariant (\ref{eq_Physical_Invariants}). Furthermore, the third expression in Eq. (\ref{eq_CFT_Left_and_Right_Central_Charges}) clearly forbids the case $c_L=c_R$, while its negative sign suggests that $c_R>c_L$. The comments above show that
\begin{equation}
0<c_L<L\,\quad L<c_R<2 \,L\,.
\end{equation}
On the other hand, from Eq. (\ref{eq_CFT_Left_and_Right_Central_Charges}), one finds the ratio of the central charges
\begin{equation}\label{eq_The_Ratio_of_the_Central_Charges}
\frac{{{c_L}}}{{{c_R}}} = \frac{{4\,{\nu ^2}}}{{3 + 5\,{\nu ^2}}}\,.
\end{equation}
It depends only on $\nu$ and certain limiting cases are valid. For $\nu\to\infty$, the ratio reaches a maximum value of $4/5$. One has to exclude this value due to  Eq. (\ref{eq_Physical_Invariants}). When $\nu=1$, the ratio is $1/2$, which is also excluded from our considerations.  Therefore,
\begin{equation}\label{eqcLcRRatioBounds}
\frac{1}{2} < \frac{{{c_L}}}{{{c_R}}} < \frac{4}{5}\,.
\end{equation}
\indent In terms of the dual CFT temperatures and charges the entropy takes the Cardy form
\begin{equation}\label{eq_Entropy_Cardy_Form}
S = \frac{{{\pi ^2}\,L}}{3}\,\left( {{c_L}\,{T_L} + {c_R}\,{T_R}} \right)\,.
\end{equation}
One can also define the following left and right moving energies,
\begin{equation}\label{eq_Left-Right_Energies}
{E_L} = \frac{{{\pi ^2}\,L}}{6}\,{c_L}\,T_L^2\,,\qquad {E_R} = \frac{{{\pi ^2}\,L}}{6}\,{c_R}\,T_R^2\,,
\end{equation}
which allow us to write the ADT conserved mass $M$ and angular momentum $J$ as
\begin{equation}\label{eq_ADT_Mass_and_Angular_Momemntum}
M = \frac{1}{G}\,\sqrt {\frac{{2\,L\,{E_L}}}{{3\,{c_L}}}} \,,\qquad J = L\,({E_L} - {E_R})\,.
\end{equation}
Consequently, the Hawking temperature $T$, defined as the surface gravity
of the horizon divided by $2\,\pi$, yields
\begin{equation}\label{eq_Hawking_Temp_via_Left-Right_Temps}
\frac{1}{T} = \frac{{4\,\pi \,\nu \,L}}{{{\nu ^2} + 3}}\,\frac{{{T_L} + {T_R}}}{{{T_R}}}\,.
\end{equation}
\indent In order to study the thermodynamic properties of the WAdS black hole, we will express the entropy $S$ in terms of the other extensive parameters $M$ and $J$, and also the ADT mass $M$ in terms of $S$ and $J$. From Eqs. (\ref{eq_Entropy_Cardy_Form})--(\ref{eq_Hawking_Temp_via_Left-Right_Temps}) one finds the explicit expressions ($G=1$):
\begin{equation}
S(M,\,J) = \pi \,\left( {M\,{c_L} + \sqrt {{c_R}\,\left( {{c_L}\,{M^2} - \frac{{2\,J}}{3}} \right)} } \right)\,,
\end{equation}
and 
\begin{equation}
M(S,\,J) = \frac{{\sqrt {3\,{c_L}\,{c_R}\,\left( {2\,{\pi ^2}\,J\,\left( {{c_R} - {c_L}} \right) + 3\,{S^2}} \right)}  - 3\,{c_L}\,S}}{{3\,\pi \,{c_L}\,\left( {{c_R} - {c_L}} \right)}}\,.
\end{equation}
It will be convenient for our considerations to also express $S$ and $M$ in terms of other parameters, namely in ($J,\,T$) space:
\begin{equation}
S(J,\,T) = \frac{{\pi \,\sqrt {2\,{c_L}\,J} \,\left( {1 + ({c_R} - {c_L})\,\pi \,T} \right)}}{{\sqrt 3 \,\sqrt {1 - {c_L}\,\pi \,T\,\left( {2+({c_R} - {c_L})\,\pi \,T} \right)} }}\,,
\end{equation}
\begin{equation}
M(J,\,T) = \frac{{\sqrt {2\,J} \,\left( {1 - {c_L}\,\pi \,T} \right)}}{{\sqrt {3\,{c_L}\,\left( {1 - {c_L}\,\pi \,T\,\left( {2 + ({c_R} - {c_L})\,\pi \,T} \right)} \right)} }}\,.
\end{equation}
We can also take advantage of the first law (\ref{eq_The_First_Law_of_TD}) to find $T$ and $\Omega$ in ($S,\,J$) space:
\begin{equation}
T(S,\,J) = \frac{{\partial M}}{{\partial S}} = \frac{1}{{\pi \,\left( {{c_L} - {c_R}} \right)}}\,\left( {1 - \frac{{\sqrt {3\,{c_R}} \,S}}{{\sqrt {{c_L}\,\left( {2\,J\,{\pi ^2}\,({c_R} - {c_L}) + 3\,{S^2}} \right)} }}} \right)\,, 
\end{equation}
\begin{equation}\label{eqOmegainS-JSpace}
\Omega (S,\,J) = \frac{{\partial M}}{{\partial J}} = \frac{{\pi \,{c_R}}}{{ \sqrt {3\,{c_L}\,{c_R}\left( {2\,J\,{\pi ^2}\,({c_R} - {c_L}) + 3\,{S^2}} \right)} }}\,.
\end{equation}
or, equivalently, in ($M,\,J$) space:
\begin{equation}
T (M,\,J)= \frac{1}{{{c_L}\,\pi }}\,{\left( {1 + \frac{{\sqrt {3\,{c_R}} \,M}}{{\sqrt { 3\,{c_L}\,{M^2}- 2\,J} }}} \right)^{ - 1}}\,,
\end{equation}
\begin{equation}\label{eqOmegainM-JSpace}
\Omega (M,\,J) = \frac{{{c_R}}}{{{c_L}\left( {3\,{c_R}\,M + \sqrt {3\,{c_R}\,\left( {3{c_L}{M^2} - 2\,J} \right)} } \right)}}\,.
\end{equation}
Naturally one requires $T>0$, which constrains the values of the left central charge:
\begin{equation}\label{eqPositiveTempCondition}
{c_L} > \frac{{2\,J}}{{3\,{M^2}}}=\frac{2\,a}{3}\,,
\end{equation}
where $a=J/M^2$ is the spin parameter. Relation (\ref{eqPositiveTempCondition}) shows that for slowly rotating system, $a<1$, the left central charge $c_L$ does not exceed 2/3. For ultra spinning black holes, $a\ge 1$, the charge $c_L$ is bounded from above only by $L$, thus we can constrain the rotation of the WAdS$_3$ solution such as
\begin{equation}\label{eqSpinBound}
a<\frac{3\,L}{2}\,.
\end{equation}
\indent In the sections below, we investigate the local and global thermodynamic properties of the WAdS$_3$ black hole solution.

\section{Heat capacity and local thermodynamic stability}\label{sec3}

One can distinguish two types of phase transitions with respect to the heat capacity $C$. The first type occurs when $C$ changes sign, i.e. $\partial_S M$=0, while the second type is identified by the singular points of the heat capacity itself, i.e. $\partial^2_{SS} M=0$. The heat capacity of the WAdS$_3$ black hole in ($S,\,J$) space is given by
\begin{equation}\label{eq_heat_capacity_in_S_J_space}
C(S,\,J) = \frac{{{\partial _S}M}}{{\partial _{SS}^2M}} = \frac{{\left( {2\,{\pi ^2}\,J\,({c_R} - {c_L}) + 3\,{{\rm{S}}^2}} \right)\,\left( {\sqrt 3 \,{{\rm{c}}_R}\,{\rm{S}} - \sqrt {{c_L}\,{c_R}\,\left( {2\,\left( {{c_R} - {c_L}} \right)\,J\,{\pi ^2} + 3\,{S^2}} \right)} } \right)}}{{2\,\sqrt 3 \,{\pi ^2}\,J\,{{\rm{c}}_R}\,({c_R} - {c_L})}}\,,
\end{equation}
where $J=0$ is a true singularity, while $c_L=c_R$ is excluded due to Eq. (\ref{eq_CFT_Left_and_Right_Central_Charges}). The heat capacity changes signs on the following spinodal curve 
\begin{equation}\label{eq_second_spinodal_curve_heat_capacity}
2\,\pi ^2\,J\,{c_L} - 3\,S^2=0\,,
\end{equation}
while $C=0$ occurs on (\ref{eq_second_spinodal_curve_heat_capacity}) or the curve
\begin{equation}\label{eqFreeQuevedo}
2\,{\pi ^2}\,J\,\left( {{c_R} - {c_L}} \right) + 3\,{S^2} = 0\,.
\end{equation}
Considering the assumptions from the previous section, these curves are not reachable by any macro state of the black hole. The condition for local thermodynamic stability, $C>0$, leads to an upper thermodynamic bound on the left central charge
\begin{equation}\label{eqLTDSCinSJSpace}
{c_L} < \frac{{3\,{S^2}}}{{2\,{\pi ^2}\,J}}.
\end{equation}
Combining Eqs. (\ref{eqLTDSCinSJSpace}) and (\ref{eqPositiveTempCondition}), one finds the following relation between the gravitational parameters:
\begin{equation}\label{eqBulkParamRelation1}
\frac{{{J^2}}}{{{M^2}\,{S^2}}} < \frac{9}{{4\,{\pi ^2}}}\,.
\end{equation}
\indent The local TD stability in ($M,\,J$) coordinates requires the explicit form of the heat capacity,
\begin{equation}\label{eq_heat_capacity_in_M_J_space}
C (M,\,J) = \frac{{\pi \,{{\rm{c}}_L}\,H\,\left( {{{\rm{c}}_R}\,\sqrt {{{\rm{c}}_R}\,\left( {3\,{{\rm{c}}_L}\,{M^2} - 2\,J} \right)}  - {{\rm{c}}_L}\,\sqrt {{{\rm{c}}_R}\,H}  + \sqrt 3 \,{{\rm{c}}_L}\,{{\rm{c}}_R}\,M} \right)}}{{2\,\sqrt 3 \,{{\rm{c}}_R}\,J\,({{\rm{c}}_R} - {{\rm{c}}_L})}}\,,
\end{equation}
where 
\begin{equation}
H = M\,\left( {2\,\sqrt {{\rm{3}}\,{{\rm{c}}_R}\,\left( {3\,{{\rm{c}}_L}\,{M^2} - 2\,J} \right)}  + 3\,M\,({{\rm{c}}_L} + {{\rm{c}}_R})} \right) - 2\,J\,.
\end{equation}
As expected, $J=0$ is a true singularity, while $C$ changes sign on
\begin{equation}
3\,M^2 \,c_L-2\,J=0\,.
\end{equation}
Imposing $C>0$, one finds agreement with (\ref{eqPositiveTempCondition}).
\\
\indent In ($J,\,T$) space the heat capacity takes the form
\begin{equation}\label{eqCinJTspace}
C(J,\,T) =T\,\frac{\partial S(J,\,T)}{\partial T}= \frac{{{c_R}\,{\pi ^2 \,T}\,\sqrt {2\,{c_L}\,J} }}{{\sqrt 3 \,{{\left( {1 - {c_L}\,\pi \,T\,\left( {2 + ({c_R} - {c_L})\,\pi \,T} \right)} \right)}^{3/2}}}}\,.
\end{equation}
When $J=0$, the specific heat $C$ changes sign, while it diverges at the following critical temperature
\begin{equation}\label{eqSpinodalT}
{T_c} = \frac{1}{{\pi \,({c_L} + \sqrt {{c_L}\,{c_R}} )}}\,.
\end{equation}
The local TD stability, $C>0$, requires
\begin{equation}\label{eqLTDSinJTspace}
T < {T_c}\,,\quad J>0\,.
\end{equation}
\indent Finally, the heat capacity in ($S,\,\Omega$) space is written by
\begin{equation}
C(S,\,\Omega)= \frac{{{{\rm{c}}_R}\,{\pi ^2}\,(\pi  - 3\,S\,\Omega )}}{{3\,\Omega \,(9\,{{\rm{c}}_L}\,{S^2}\,{\Omega ^2} - {\pi ^2}\,{{\rm{c}}_R})}}\,.
\end{equation}
It changes sign at 
\begin{equation}\label{eqHeatCapSignChangeOmega-SSpace}
\Omega\,S=\frac{\pi}{3}\,,
\end{equation}
while its divergences are located on
\begin{equation}\label{eqHeatCapDivOmega-SSpace}
{\Omega ^2}\,{S^2} = \frac{{{{\rm{c}}_R}\,{\pi ^2}}}{{9\,{{\rm{c}}_L}}}\,.
\end{equation}
The singularity at $\Omega=0$ is excluded due to Eq. (\ref{eqTandOmegainRSpace}).
The local TD stability in ($S,\,\Omega$) space defines the region
\begin{equation}\label{eqOmega-SLocalTDStability}
\frac{\pi }{3} < \Omega \,S < \frac{\pi }{3}\,\sqrt {\frac{{{c_R}}}{{{c_L}}}} \,.
\end{equation}
\indent Further constraints on the CFT central charges or the gravitational parameters can be found by considering the methods of thermodynamic information geometry, as shown in the following sections. Also, more on the thermodynamic stability of the WAdS$_3$, in the space of the original gravitational parameters, can be found in \cite{Birmingham:2010mj}.

\section{Hessian thermodynamic information metrics}\label{sec4}

\subsection{Ruppeiner metric}
We begin by calculating the Ruppeiner thermodynamic metric given by the Hessian of the entropy
\begin{equation}
g_{ab}^{(R)}=-\partial_a\partial_b S(M,\,J)\,,\quad a,\,b=(M,\,J)\,.
\end{equation}
The explicit form of the metric is written by
\begin{equation}\label{eq_Ruppeiner_metric_Explicit}
{{\hat g}^{(R)}} = \left( {\begin{array}{*{20}{c}}
{g_{MM}^{(R)}}&{g_{MJ}^{(R)}}\\
{g_{JM}^{(R)}}&{g_{JJ}^{(R)}}
\end{array}} \right) = \left( {\begin{array}{*{20}{c}}
{\frac{{2\,\pi \,J\,{{\rm{c}}_L}\,\sqrt {3\,{c_R}\,\left( {3\,{c_L}\,{M^2} - 2\,J} \right)} }}{{{{\left( {3\,{c_L}\,{M^2} - 2\,J} \right)}^2}}}}&{ - \frac{{\sqrt 3 \,\pi \,M\,{{\rm{c}}_L}{\rm{c}}_R^2}}{{\sqrt {{{\left( {{c_R}\,\left( {3\,{c_L}\,{M^2} - 2\,J} \right)} \right)}^3}} }}}\\
{ - \frac{{\sqrt 3 \,\pi \,M\,{{\rm{c}}_L}{\rm{c}}_R^2}}{{\sqrt {{{\left( {{c_R}\,\left( {3\,{c_L}\,{M^2} - 2\,J} \right)} \right)}^3}} }}}&{\frac{{\pi \,{\rm{c}}_R^2}}{{\sqrt 3 \,\sqrt {{{\left( {{c_R}\,\left( {3\,{c_L}\,{M^2} - 2\,J} \right)} \right)}^3}} }}}
\end{array}} \right)\,.
\end{equation}
\indent In order to identify any critical points and phase transitions we investigate the singularities of the thermodynamic curvature with respect to the metric (\ref{eq_Ruppeiner_metric_Explicit}),
\begin{equation}
R_I^{(R)} =  - \frac{{\sqrt 3 }}{{\pi \,\sqrt {{c_R}\,\left( {3\,{c_L}\,{M^2} - 2\,J} \right)} }}\,.
\end{equation}
This expression shows that $R_I^{(R)}$ is singular at $c_L=2\,J/(3\,M^2)$, but  regular for $J=0$. Therefore, the Ruppeiner information metric accounts only for half of the necessary critical points of the heat capacity (\ref{eq_heat_capacity_in_M_J_space}) and one can not use it to fully describe the equilibrium space of the WAdS$_3$ black hole solution.

\subsection{Weinhold metric}
On the other hand, one can consider the Weinhold information metric,
\begin{equation}
g_{ab}^{(W)}=\partial_i\partial_j M(S,\,J)\,,\quad i,\,j=(S,\,J)
\end{equation}
with components
\begin{equation}
{{\hat g}^{(W)}} = \left( {\begin{array}{*{20}{c}}
{g_{SS}^{(W)}}&{g_{SJ}^{(W)}}\\
{g_{JS}^{(W)}}&{g_{JJ}^{(W)}}
\end{array}} \right) = \left( {\begin{array}{*{20}{c}}
{\frac{{2\,\sqrt 3 \,J\,\pi \,{{\rm{c}}_L}\,c_R^2}}{{{h^{3/2}}}}}&{ - \frac{{\sqrt 3 \,\pi \,{{\rm{c}}_L}\,c_R^2\,S}}{{{h^{3/2}}}}}\\
{ - \frac{{\sqrt 3 \,\pi \,{{\rm{c}}_L}\,c_R^2\,S}}{{{h^{3/2}}}}}&{ - \frac{{{\pi ^3}\,{{\rm{c}}_L}\,c_R^2\,({{\rm{c}}_R} - {{\rm{c}}_L})}}{{\sqrt 3 \,{h^{3/2}}}}}
\end{array}} \right)\,,
\end{equation}
where $h = {{\rm{c}}_L}\,{{\rm{c}}_R}\,\left( {3\,{S^2} + 2\,({{\rm{c}}_R} - {{\rm{c}}_L})\,J\,{\pi ^2}} \right)$. 
In this case, the Weinhold curvature,
\begin{equation}
R_I^{(W)} = \frac{{\sqrt 3 \,\pi \,{{\rm{c}}_L}\,({{\rm{c}}_R} - {{\rm{c}}_L})}}{{\sqrt {{{\rm{c}}_L}\,{{\rm{c}}_R}\,\left( {2\,{\pi ^2}\,J\,({{\rm{c}}_R} - {{\rm{c}}_L}) + 3\,{S^2}} \right)} }}\,,
\end{equation}
does not reproduce the singularities of the heat capacity (\ref{eq_heat_capacity_in_S_J_space}) in ($S,\,J$) space. Therefore, it is also not a suitable Riemannian metric on the equilibrium state space of the WAdS$_3$ black hole.

\subsection{Ensembles and global thermodynamic stability}

By choosing other potentials, one can construct Hessian thermodynamic metrics, which, in general, are not conformally related to Ruppeiner or Weinhold metrics. For example, in the canonical ensemble we can consider the Helmholtz potential,
\begin{equation}
F(M,\,J) = M - T\,S = \frac{{2\,J\,\sqrt {{c_R}} }}{{{c_L}\,\left( {3\,M\,\sqrt {{c_R}}  + \sqrt 3 \,\sqrt {3\,{c_L}\,{M^2} - 2\,J} } \right)}}\,,
\end{equation}
which defines the following Hessian thermodynamic metric:
\begin{align}
&g_{MM}^{(H)} = \frac{{36\,J\,\sqrt {{c_R}} \,\left( {3\,{c_L}\,{M^2}\,\left( {\left( {{c_L} + {c_R}} \right)\,f + 2\,{c_L}\,M\,\sqrt {3\,{c_R}} } \right) + J\,\left( {{c_L}\,\left( {f - 3\,M\,\sqrt {3\,{c_R}} } \right) - 2\,{c_R}\,f} \right)} \right)}}{{{c_L}\,{{\left( {\sqrt 3 \,f + 3\,M\,\sqrt {{c_R}} } \right)}^3}\,{f^3}}}\,,
\\
&g_{MJ}^{(H)} = \frac{{18\,M\,\sqrt {{c_R}} \,\left( {3\,{c_L}\,J\,M\,\sqrt {3\,{c_R}}  - \left( {{c_L} - 2\,{c_R}} \right)\,f\,J - 3\,{c_L}\,\left( {{c_L} + {c_R}} \right)\,f\,{M^2} - 6\,c_L^2\,{M^3}\,\sqrt {3\,{c_R}} } \right)}}{{{c_L}\,{f^3}\,{{\left( {\sqrt 3 \,f + 3\,M\,\sqrt {{c_R}} } \right)}^3}}}\,,
\\
&g_{JJ}^{(H)} =  - \frac{{6\,\left( {f\,\sqrt {{c_R}} \,\left( {J - 6\,{c_L}\,{M^2}} \right) + 3\,\sqrt 3 \,{c_R}\,M\,\left( {J - 2\,{c_L}\,{M^2}} \right)} \right)}}{{{c_L}\,{f^3}\,{{\left( {\sqrt 3 \,f + 3\,M\,\sqrt {{c_R}} } \right)}^3}}}\,,
\end{align}
where $g^{(H)}_{ij}=\partial_i\partial_j F(M, J)$ and $f = \sqrt {3\,{c_L}\,{M^2} - 2\,J} $. The denominator of the Helmholtz thermodynamic curvature,
\begin{align}
\nonumber {\rm{denom}}\,(R_I^{(H)}) &=  - 2\,{f^2}\,{\left( {\left( {{c_L} - 2\,{c_R}} \right)\,f\,J - 3\,{c_L}\,J\,M\,\sqrt {3\,{c_R}}  + 3\,{c_L}\,\left( {{c_L} + {c_R}} \right)\,f\,{M^2} + 6\,c_L^2\,{M^3}\,\sqrt {3\,{c_R}} } \right)^2}\\
 &\times {\left( {\sqrt {{c_R}} \,f\,\left( {J - 3\,{c_L}\,{M^2}} \right) + 3\,\sqrt 3 \,{c_R}\,M\,\left( {J - 2\,{c_L}\,{M^2}} \right) - 3\,c_R^{3/2}\,f\,{M^2}} \right)^2}\,,
\end{align}
is zero, when $c_L=2\,J/(3\,M^2)$, but non-zero for $J=0$. The numerator of the TD curvature is also finite, when $J\to 0$. 
\\
\indent The situation does not improve in the grand canonical ensemble as well, where one works with the Gibbs free energy
\begin{equation}
G(M,\,J) = M - T\,S - \Omega \,J = \frac{{J\,\sqrt {3\,{c_R}} }}{{{c_L}\,\left( {M\,\sqrt {3\,{c_R}}  + \sqrt {3\,{c_L}\,{M^2} - 2\,J} } \right)}}\,.
\end{equation}
Therefore, we can make the conclusion that Hessian thermodynamic geometries are not well-suited for the description of the WAdS$_3$ BH equilibrium space.
\\
\indent Before addressing the problem of global thermodynamic stability of the warped AdS black hole, we need to write the explicit expressions for the Helmholtz and the Gibbs free energies in ($J,\,T$) space:
\begin{equation}\label{eqHelmholtzinJTSpace}
F(J,\,T) = \frac{{\sqrt {2\,J} }}{{\sqrt {3\,{c_L}} }}\,\sqrt {1 - {c_L}\,\pi \,T\,\left( {2 + \left( {{c_L} - {c_R}} \right)\,\pi \,T} \right)} \,,
\end{equation}
\begin{equation}
G(J,\,T) = \frac{{\sqrt J }}{{\sqrt {2\,{c_L}} }}\,\sqrt {1 - {c_L}\,\pi \,T\,\left( {2 + ({c_R} - {c_L})\,\pi \,T} \right)} \,.
\end{equation}
The conditions for global TD stability in the canonical and grand canonical ensembles, respectively, are given by \cite{LUBARDA200848}:
\begin{equation}\label{eqConcavityofFandG}
\frac{{\partial^2 F}}{{\partial T^2}} < 0,\qquad \frac{{\partial^2 G}}{{\partial T^2}} < 0\,.
\end{equation}
Here, the concavity of the Helmholtz free energy $F$ and the Gibbs energy $G$, with respect to the temperature $T$, implies that the corresponding specific heats are both positive. Both conditions (\ref{eqConcavityofFandG}) are satisfied for $T<T_c$, which is also the requirement (\ref{eqLTDSinJTspace}) for local TD stability. Hence, the WAdS$_3$ black hole solution is locally and globally stable, from thermodynamic standpoint, in the same temperature interval $0<T<T_c$, whereas unstable for $T>T_c$. Near the critical temperature, $T=T_c$, the underlying inter-particle interactions becomes strongly correlated and phase transitions occur. In this case, the equilibrium description of the black hole system breaks down.
\section{Legendre invariant information metrics}\label{sec5}

\subsection{Quevedo metrics}

The Quevedo information metric on the equilibrium state space of the WAdS$_3$ solution is given by \cite{Quevedo:2017tgz}:
\begin{equation}
ds_Q^2 = W\,( - \partial _S^2M\,d{S^2} + \partial _J^2M\,d{J^2}) = g_{SS}^{(Q)}\,d{S^2} + g_{JJ}^{(Q)}\,d{J^2}\,,
\end{equation}
where the conformal function $W$ has one of the following forms
\begin{equation}
W = \left\{ \begin{array}{l}
S\,\frac{{\partial M}}{{\partial S}} + J\,\frac{{\partial M}}{{\partial J}}\,,\quad \text{case\,I}\,,\\
\\
S\,\frac{{\partial M}}{{\partial S}}\,,\quad \text{case\,II}\,.
\end{array} \right.
\end{equation}
The first expression for $W$ leads to the following components of the information metric:
\begin{align}
&g_{SS}^{(Q,\, I)} = \frac{{2\,J\,c_R^2\,{c_L}\,\left( {\sqrt 3 \,{{\rm{c}}_R}\,\left( {{\pi ^2}\,J\,({{\rm{c}}_R} - {{\rm{c}}_L}) + 3\,{S^2}} \right) - 3\,S\,\sqrt h } \right)}}{{\sqrt 3 \,({{\rm{c}}_L} - {{\rm{c}}_R})\,{h^2}}}\,,\\
&g_{JJ}^{(Q,\,I)} = \frac{{{\pi ^2}\,c_R^2\,{c_L}\,\left( {{{\rm{c}}_R}\,\left( {{\pi ^2}\,J\,({{\rm{c}}_L} - {{\rm{c}}_R}) - 3\,{S^2}} \right) + S\,\sqrt {3\,h} } \right)}}{{3\,{{\rm{c}}_L}\,{h^2}}}\,,
\end{align}
where  $h = {{\rm{c}}_L}\,{{\rm{c}}_R}\,\left( {3\,{S^2} + 2\,({{\rm{c}}_R} - {{\rm{c}}_L})\,J\,{\pi ^2}} \right)$. 
The denominator of the thermodynamic curvature for case I reads
\begin{align}
{\rm{denom}}(R_I^{(Q,\,I)}) = 6\,{J^2}\,{\pi ^2}\,{\left( {S\,(\sqrt {3\,h}  - 3\,{{\rm{c}}_R}\,S) - {\pi ^2}\,{{\rm{c}}_R}\,J\,({{\rm{c}}_R} - {c_L})} \right)^4}\,.
\end{align}
It is singular at $J=0$. One notes that the curve (\ref{eq_second_spinodal_curve_heat_capacity}) is not reproduced by ${\rm{denom}}(R_I^{(Q,\,I)})$. However, we excluded this curve as a true singularity of the heat capacity. Therefore, we can consider this metric as viable TD metric, if we treat $J=0$ as a first-order phase transition point.
\\
\indent On the other hand, the metric components in case II are given by
\begin{equation}
\hat g_{}^{(Q,\, II)} = \left( {\begin{array}{*{20}{c}}
{g_{SS}^{(Q,{\kern 1pt} II)}}&0\\
0&{g_{JJ}^{(Q,{\kern 1pt} II)}}
\end{array}} \right) = \left( {\begin{array}{*{20}{c}}
{\frac{{2{c_L}c_R^2\,J\,S\,\left( {3\,{{\rm{c}}_R}\,S - \sqrt {3\,h} } \right)}}{{({c_L} - {{\rm{c}}_R})\,{h^2}}}}&0\\
0&{\frac{{{c_L}\,c_R^2\,{\pi ^2}\,S\,\left( {\sqrt {3\,h}  - {3_R}\,S} \right)}}{{3\,{h^2}}}}
\end{array}} \right)\,.
\end{equation}
The denominator of the second Quevedo curvature, 
\begin{equation}\label{eq_secon_Q_curvature}
{\rm{denom}}(R_I^{(Q,\,II)}) = 6\,\sqrt 3 \,{\pi ^2}\,{J^2}\,{S^3}\,{\left( {2\,{\pi ^2}\,{{\rm{c}}_L}\,J\,({{\rm{c}}_R} - {c_L}) + 3\,{S^2}\,({{\rm{c}}_L} + {{\rm{c}}_R}) - 2\,\sqrt {3\,h} \,S} \right)^3}\,,
\end{equation}
is singular at $J=0$. It is also singular at the spinodal curve (\ref{eq_second_spinodal_curve_heat_capacity}) of the heat capacity, which we discarded. As discussed in Section \ref{sec2}, the case $S=0$ is excluded, thus $R_I^{(Q,II)}$ does not show additional singularities. Therefore, the second Quevedo information metric is a viable thermodynamic metric.
\\
\indent We can now consider the values of the $R_I^{(Q,II)}$. In the region of TD stability, $R_I^{(Q,II)}$ is negative, which defines a hyperbolic thermodynamic information geometry on the equilibrium manifold. According to the established interpretation, the negative values of the TD curvature correspond to attractive inter-particle interactions. The free, non-interacting quantum theory, $R_I^{(Q,II)}=0$, occurs on the level curve $h=0$, i.e.
\begin{equation}\label{}
2\,{\pi ^2}\,J\,\left( {{c_R} - {c_L}} \right) + 3\,{S^2} = 0\,.
\end{equation}
Under the current assumptions, ($J\geq 0,\,S>0,\,c_R>c_L$), this curve is not reachable by the macro states of the black hole. Therefore, in Quevedo's approach, we always have an interacting theory.

\subsection{HPEM metric}

As an attempt to avoid extra singular points in the Quevedo thermodynamic curvature, which do not coincide with phase transitions of any type, in \cite{Hendi:2015rja} the authors proposed an alternative information metric with different conformal factor,
\begin{equation}
ds_{HPEM}^2 = S\,\frac{{{\partial _S}M}}{{{{(\partial _J ^2M)}^3}}}\,( - \partial _S^2M\,d{S^2} + \partial _J ^2M\,d{J ^2})\,.
\end{equation}
Its explicit components are given by
\begin{align}
{{\hat g}^{(HPEM)}} = \left( {\begin{array}{*{20}{c}}
{g_{SS}^{(HPEM)}}&0\\
0&{g_{JJ}^{(HPEM)}}
\end{array}} \right) = \left( {\begin{array}{*{20}{c}}
{\frac{{18\,{h^{5/2}}\,J\,S\,\left( {\sqrt 3 \,{{\rm{c}}_R}\,S - \sqrt h } \right)}}{{{\pi ^9}\,{\rm{c}}_L^2\,{{({{\rm{c}}_L} - {{\rm{c}}_R})}^4}\,{\rm{c}}_R^4}}}&0\\
0&{\frac{{3\,{h^{5/2}}\,S\,\left( {\sqrt h  - \sqrt 3 \,{{\rm{c}}_R}\,S} \right)}}{{{\pi ^7}\,{\rm{c}}_L^2\,{{({{\rm{c}}_L} - {{\rm{c}}_R})}^3}\,{\rm{c}}_R^4}}}
\end{array}} \right)\,.
\end{align}
In this case, the denominator of the HPEM curvature is given by
\begin{equation}\label{eqHPEMTDcurvatureDenom}
{\mathop{\rm denom}\nolimits} (R_I^{(HPEM)}) = 18\,{J^2}\,{S^3}\,{\left( {\sqrt h  - \sqrt 3 \,{c_R}\,S} \right)^3}\,{h^{7/2}}\,.
\end{equation}
It covers all relevant divergences of the heat capacity (\ref{eq_heat_capacity_in_S_J_space}) including the curve (\ref{eq_second_spinodal_curve_heat_capacity}), which we excluded. At first, it seems that a new spinodal curve $h=0$ is present in (\ref{eqHPEMTDcurvatureDenom}), but, as mentioned already, under the assumptions ($J\geq 0,\,S>0,\,c_R>c_L$), the macro states can not lie on this curve. Therefore, the HPEM curvature is healthy from additional relevant singularities.
\\
\indent Contrary to Quevedo's case, here, one can find the parameter region, where the Sylvester criterion holds together with $C>0$. It coincides with the region (\ref{eqLTDSCinSJSpace}) for TD stability in ($S,\,J$) space. Furthermore, in this region, $R_I^{HPEM}$ is positive, which defines an elliptic thermodynamic information geometry on the equilibrium manifold. The latter corresponds to repulsive inter-particle interactions.

\section{Information metrics and conjugate thermodynamic potentials}\label{sec6}

\subsection{MM Metric I}
Another geometric approach to the problem of equilibrium state space of black holes was proposed by A. H. Mansoori
and B. Mirza in \cite{Mansoori:2013pna}. The authors define a conjugate thermodynamic potential as
an appropriate Legendre transformation of the thermal parameters in order to mach the divergences
of the specific heat.
For example, one can consider the following conjugate potential
\begin{equation}
K(S,\Omega ) = M(S,\Omega ) - \Omega \,J(S,\Omega )=\frac{{{c_R}\,{\pi ^2} + 3\,{c_L}\,S\,\Omega \,\left( {3\,S\,\Omega  - 2\,\pi } \right)}}{{6\,{c_L}\,{\pi ^2}\,\Omega \,\left( {{c_R} - {c_L}} \right)}}\,,
\end{equation}
where
\begin{equation}
M(S,\Omega ) = \frac{{{c_R}\,\pi  - 3\,{c_L}S\,\Omega }}{{3\,{c_L}\,\pi \,\Omega \,\left( {{c_R} - {c_L}} \right)}}\,,\qquad J(S,\Omega ) = \frac{{{c_R}\,{\pi ^2} - 9\,{c_L}{S^2}\,{\Omega ^2}}}{{6\,{c_L}\,{\pi ^2}\,{\Omega ^2}\,\left( {{c_R} - {c_L}} \right)}}\,.
\end{equation}
In this case, the components of the information metric read:
\begin{align}
&g_{SS}^{(MM)} = \frac{1}{T}\,\frac{{{\partial ^2}K}}{{\partial {S^2}}} = \frac{{3\,\Omega }}{{3\,S\,\Omega  - \pi }}\,,\\
&g_{S\Omega }^{(MM)} = \frac{1}{T}\,\frac{{{\partial ^2}K}}{{\partial \Omega \,\partial S}} = \frac{{3\,S}}{{3\,S\,\Omega  - \pi }}\,,\\
&g_{\Omega \Omega }^{(MM)} = \frac{1}{T}\,\frac{{{\partial ^2}K}}{{\partial {\Omega ^2}}} = \frac{{{c_R}\,{\pi ^2}}}{{3\,{c_L}\,{\Omega ^3}\,(3\,S\,\Omega  - \pi )}}\,,
\end{align}
where the temperature in ($S,\,\Omega$) space is written by
\begin{equation}
T(S,\Omega ) = \frac{{3\,S\,\Omega  - \pi }}{{\left( {{c_R} - {c_L}} \right)\,{\pi ^2}}}\,.
\end{equation}
Under the set of conditions $\{S>0,\,\Omega>0\}$, together with the conditions for the central charges from Section \ref{sec3}, the temperature is positive, if
\begin{equation}\label{eqSOmegaLocalTDstability}
\Omega \,S > \frac{\pi }{3}\,,
\end{equation}
whereas the angular momentum is non-negative, if
\begin{equation}
\Omega \,S \le \frac{\pi }{3}\,\sqrt {\frac{{{c_R}}}{{{c_L}}}} \,.
\end{equation}
On the other hand $C>0$ forbids the equal sign, thus we arrive at the condition for the local TD stability (\ref{eqOmega-SLocalTDStability}). The MM thermodynamic curvature yields
\begin{align}
R_I^{(MM,I)} = \frac{{3\,\pi \,\Omega \,\left( {{c_L}\,{c_R}\,\pi \,\left( {9\,\pi \,S\,\Omega  - 27\,{S^2}\,{\Omega ^2} - 2\,{\pi ^2}} \right) - c_R^2\,{\pi ^3} - 27\,c_L^2\,{S^3}\,{\Omega ^3}} \right)}}{{\left( {3\,S\,\Omega  - \pi } \right)\,{{\left( {{c_R}\,{\pi ^2} - 9\,{c_L}{S^2}\,{\Omega ^2}} \right)}^2}}}\,.
\end{align}
It exactly matches the spinodal curves (\ref{eqHeatCapSignChangeOmega-SSpace}) and (\ref{eqHeatCapDivOmega-SSpace}) of the heat capacity. Thus the proposed metric is a viable choice for the description of the equilibrium space of the WAdS black hole.
\\
\indent The compact expression for $R_I^{(MM,I)}$ allows us to find the conditions for its sign. For example, the MM curvature is positive, if
\begin{equation}
\frac{{{c_L}}}{{{c_R}}} > \frac{{\pi \,\left( {2\,{\pi ^2} - 9\,\pi \,S\,\Omega  + 27\,{S^2}\,{\Omega ^2} + \left( {\pi  - 3\,S\,\Omega } \right)\,\sqrt {4\,{\pi ^2} - 12\,\pi \,S\,\Omega  + 81\,{S^2}\,{\Omega ^2}} } \right)}}{{54\,{S^3}\,{\Omega ^3}}}\,.
\end{equation}
On the other hand, $R_I^{(MM,I)}$ can become negative, if
\begin{equation}\label{eqMMRnegative1}
\frac{{{c_L}}}{{{c_R}}} < \frac{{\pi \,\left( {2\,{\pi ^2} - 9\,\pi \,S\,\Omega  + 27\,{S^2}\,{\Omega ^2} + \left( {\pi  - 3\,S\,\Omega } \right)\,\sqrt {4\,{\pi ^2} - 12\,\pi \,S\,\Omega  + 81\,{S^2}\,{\Omega ^2}} } \right)}}{{54\,{S^3}\,{\Omega ^3}}}\,,
\end{equation}
or
\begin{equation}\label{eqMMRnegative2}
\frac{{{c_L}}}{{{c_R}}} > \frac{{\pi \,\left( {2\,{\pi ^2} - 9\,\pi \,S\,\Omega  + 27\,{S^2}\,{\Omega ^2} - \left( {\pi  - 3\,S\,\Omega } \right)\,\sqrt {4\,{\pi ^2} - 12\,\pi \,S\,\Omega  + 81\,{S^2}\,{\Omega ^2}} } \right)}}{{54\,{S^3}\,{\Omega ^3}}}\,.
\end{equation}
The MM curvature is zero, if expressions (\ref{eqMMRnegative1}) and (\ref{eqMMRnegative2}) are treated as equations separately. In any case, the values of the right hand side of both expressions should respect Eq. (\ref{eqcLcRRatioBounds}).
\\
\indent Finally, we also find that Sylvester's criterion is compatible with the condition (\ref{eqOmega-SLocalTDStability}) for local TD stability.

\subsection{MM Metric II} 
Another proposal for information metric by MM involves the Helmholtz potential in ($J, \,T$) space (\ref{eqHelmholtzinJTSpace}). The new information metric is given by
\begin{align}
&g_{TT}^{(MM)} = \frac{1}{T}\,\frac{{{\partial ^2}F}}{{\partial {T^2}}} =  - \frac{{{c_R}\,{\pi ^2}\,\sqrt {2\,{c_L}\,J} }}{{\sqrt 3 \,T\,{{\left( {1 - {c_L}\,\pi \,T\,\left( {2 + ({c_R} - {c_L})\,\pi \,T} \right)} \right)}^{3/2}}}}\,,\\
&g_{TJ}^{(MM)} = \frac{1}{T}\,\frac{{{\partial ^2}F}}{{\partial T\,\partial J}} =  - \frac{{{c_L}\,\pi \,\left( {1 + ({c_R} - {c_L})\,\pi \,T} \right)}}{{\sqrt 6 \,T\,\sqrt {{c_L}\,J\,\left( {1 - {c_L}\,\pi \,T\,\left( {2 + \left( {{c_R} - {c_L}} \right)\,\pi \,T} \right)} \right)} }}\,,\\
&g_{JJ}^{(MM)} = \frac{1}{T}\,\frac{{{\partial ^2}F}}{{\partial {J^2}}} =  - \frac{{\sqrt {{c_L}\,J\,\left( {1 - {c_L}\,\pi \,T\,\left( {2 + \left( {{c_R} - {c_L}} \right)\,\pi \,T} \right)} \right)} }}{{2\,\sqrt 6 \,{c_L}\,{J^2}\,T}}\,.
\end{align}
Now, the MM thermodynamic curvature yields
\begin{align}
R_I^{(MM,II)} = \frac{{\sqrt 3 \,\left( {{c_L}\,\pi \,T\,\left( {3 - \left( {{c_R} - {c_L}} \right)\,\left( {{c_L}\,\pi \,T - 3} \right)\,\pi \,T} \right) - 1} \right)}}{{{\pi ^2}\,T\,\sqrt {2\,{c_L}\,J} \,\left( {{c_L} - {c_R}} \right)\,{{\left( {1 - {c_L}\,\pi \,T\,\left( {2 + \left( {{c_R} - {c_L}} \right)\,\pi \,T} \right)} \right)}^{3/2}}}}\,.
\end{align}
It respects all relevant spinodals of the heat capacity (\ref{eqCinJTspace}) in ($J, \,T$) space. Moreover, one notes that both MM proposals has no redundant singularities in the corresponding TD curvatures. Here, we also note that the Sylvester criterion is incompatible with the condition for local thermodynamic stability.
\\
\indent The sign of the second MM curvature is as follow. First, one has $R_I^{(MM,II)}>0$, if
\begin{equation}
{c_L}\,\pi \,T\,\left( {3 - \left( {{c_R} - {c_L}} \right)\,\left( {{c_L}\,\pi \,T - 3} \right)\,\pi \,T} \right) < 1\,.
\end{equation}
Consequently, one has $R_I^{(MM,II)}<0$ in several regions. The first one is given by
\begin{equation}
3\,{c_L}\,\pi \,T \leq 1,\qquad {c_L}\,\pi \,T\,\left( {3 - \left( {{c_R} - {c_L}} \right)\,\left( {{c_L}\,\pi \,T - 3} \right)\,\pi \,T} \right) > 1\,.
\end{equation}
The second one is
\begin{equation}
3\,{c_R}\,\pi \,T < 1\,.
\end{equation}
And finally, the third region is
\begin{equation}
\frac{1}{{3\,\pi \,{c_L}}} < T < \frac{1}{{2\,\pi \,{c_L}}}\,.
\end{equation}
A flat information geometry, $R_I^{(MM,II)}=0$, occurs on the following level curve
\begin{equation}\label{eqMMFlatInfoGeomII}
{{c_L}\,\pi \,T\,\left( {3 - \left( {{c_R} - {c_L}} \right)\,\left( {{c_L}\,\pi \,T - 3} \right)\,\pi \,T} \right) = 1}\,.
\end{equation}
In any cases, all inequalities for the sign of the curvature should respect the condition for global/local TD stability, $T<T_c$.

\section{Complexity growth of the WAdS$_3$ black hole}\label{sec7}

In this section we compute the complexity growth rate of the WAdS$_3$ black hole solution (\ref{eq_WAdS3_Metric}) within the ``Complexity=Action'' conjecture\footnote{The complexity growth rate of the warped AdS$_3$ black hole was first obtained in \cite{Ghodrati:2017roz} using the first order formalism.  Unfortunately, we suspect that the result (2.28) in \cite{Ghodrati:2017roz} is incorrect due to an error in the computation  of the Einstein-Hilbert part of the on-shell action. The correct formula for this term should be (see Eq. (2.25) in \cite{Ghodrati:2017roz}):
\[{\epsilon _{ABC}}{R^{AB}}\wedge {e^C} = -3 \ell dt \wedge dr \wedge d\theta ,
\]
which coincides with the on-shell part $\sqrt{-g}R=-3\ell$. For consitency of notations we use $\ell=L$ and we have reintroduced the $\wedge$ product. One also has $\epsilon^{t r\theta}=+1$, and the curvature 2-form $R^{AB}$ defined by
\[
{R^{AB}} = d{\omega ^{AB}} + \omega _{\,\,C}^A\wedge{\omega ^{CB}}.
\]
}. Consequently, we use the generalizations to the Lloyd bound of the rate of complexity, proposed by various authors in \cite{Cai:2016xho, Huang:2016fks}, to find further non-trivial restrictions on the equilibrium state space parameters of the system. 

The ``Complexity=Volume'' and the ``Complexity=Action'' conjectures for the WAdS$_3$ solution in Einstein's theory was analyzed in \cite{Auzzi:2018zdu} and \cite{Auzzi:2018pbc}, respectively. Similar studies for the BTZ black hole were conducted in \cite{Cai:2016xho, Ghodrati:2017roz, HosseiniMansoori:2017tsm, Qaemmaqami:2017lzs}. Further examples, regarding different aspects on the subject, can also be found in \cite{Bhattacharya:2018oeq, Carmi:2017jqz, Nagasaki:2017kqe, Tanhayi:2018gcj, HosseiniMansoori:2018gdu, Cano:2018aqi, Chapman:2018lsv, Chapman:2018dem, Carmi:2016wjl, Mazhari:2016yng, Momeni:2016ekm, Momeni:2017mmc}.
\\
\indent In order to compute the complexity growth rate of the WAdS$_3$ solution from Eq. (\ref{eq_WAdS3_Metric}) we have to vary the TMG action (\ref{eq_TMG_action}) in the WDW patch:
\begin{equation}
\delta I = I{[t + \delta t]_{WDW}} - I{[t]_{WDW}}\,.
\end{equation}
First, we take the variation of the Einstein-Hilbert term:
\begin{align}
\delta {I_{EH}} = \frac{1}{{16\,\pi }}\,\int\limits_0^{2\,\pi}  {\int\limits_{{r_ - }}^{{r_ + }} {\int\limits_t^{t + \delta t} {\sqrt { - g} \,\left( {R + \frac{2}{{{L^2}}}} \right)} } } \,dt\,dr\,d\theta  
=  - \frac{{L\,({r_ + } - {r_ - })}}{4}\,\delta t\,.
\end{align}
Next, we vary the CS term:
\begin{equation}
\delta {I_{CS}} = \frac{L}{{96\,\pi \,\nu }}\,\int\limits_0^{2\,\pi}  {\int\limits_{{r_ - }}^{{r_ + }} {\int\limits_t^{t + \delta t} {3\,\nu \,({\nu ^2} - 1)} } } \,dt\,dr\,d\theta  = \frac{{L\,({\nu ^2} - 1)\,({r_ + } - {r_ - })}}{{16}}\,\delta t\,.
\end{equation}

To calculate the contribution coming from the boundary term $B$ one has to choose a tetrad co-frame $e^A=e_{\mu}^{\,\,A} dx^\mu$ of the metric $g_{\mu\nu}=e_{\mu}^{\,\,A} e_{\nu}^{\,\,B} \eta_{AB}$, $\eta={\rm{diag}}(-1,1,1)$, namely \cite{Chen:2009hg}:
\begin{align}
\nonumber
&{e^0} = \frac{L}{{2\sqrt {D(r)} }}d\theta  = e_{\theta}^{\,\,0}d\theta ,\quad {e^1} = L\sqrt {D(r)} dr = e_{r}^{\,\,1}dr,\\
&{e^2} = Ldt + LM(r)d\theta  = e_{t}^{\,\,2}dt + e_{\theta}^{\,\,2}d\theta .
\end{align}
It is easy to verify that $g_{\mu\nu}=e_{\mu}^{\,\,A} e_{\nu}^{\,\,B} \eta_{AB}$ holds. The spin connections, ${\omega ^{AB}} = \omega_{\mu }^{\,\,AB}d{x^\mu }$, can be calculated explicitly from the tetrad co-frame by
\begin{equation}
\omega _\mu ^{\,\,AB} = \frac{1}{2}{e^{\nu A}}({\partial _\mu }e_\nu ^{\,\,B} - {\partial _\nu }e_\mu ^{\,\,B}) - \frac{1}{2}{e^{\nu B}}({\partial _\mu }e_\nu ^{\,\,A} - {\partial _\nu }e_\mu ^{\,\,A}) - \frac{1}{2}{e^{\rho A}}{e^{\sigma B}}({\partial _\rho }e_{\sigma C}^{} - {\partial _\sigma }e_{\rho C}^{})e_\mu ^{\,\,C},
\end{equation}
where ${e^{\nu A}} = {g^{\nu \lambda }}e_\lambda ^{\,\,A}$ and $e_{\sigma C}^{} = e_\sigma ^{\,\,F}{\eta _{FC}}$. We find
\begin{align}
\nonumber
&\omega _t^{\,\,01} =  - \omega _t^{\,\,10} =  - M',\quad \omega _r^{\,\,02} =  - \omega _r^{\,\,20} =  - \sqrt D M',\\
&\omega _\theta ^{\,\,01} =  - \omega _\theta ^{\,\,10} = MM' - N',\quad \omega _\theta ^{\,\,12} =  - \omega _\theta ^{\,\,21} =  - \frac{{M'}}{{2\sqrt D }}.
\end{align}
Therefore, the boundary 2-form computes to
\begin{equation}
B =  - \frac{{LD'}}{{64\pi {D^2}}}dt \wedge d\theta  = \frac{{L({\nu ^2} + 3)(2r - {r_ - } - {r_ + })}}{{64\pi }}dt \wedge d\theta .
\end{equation}
Considering the above setup, the contribution from the boundary term yields
\begin{equation}
\delta {I_B} = \frac{1}{{32}}\int\limits_t^{t + \delta t} {L({\nu ^2} + 3)\left( {(2r - {r_ - } - {r_ + })\left| {_{{r_ + }}} \right. - (2r - {r_ - } - {r_ + })\left| {_{{r_ - }}} \right.} \right)} dt = \frac{{L({\nu ^2} + 3)({r_ + } - {r_ - })}}{{16}}\delta t.
\end{equation}
Finally, the result for the rate of complexity growth is
\begin{equation}
\dot C = \frac{L}{{8}}\,({\nu ^2}- 1)\,({r_ + } - {r_ - }) = \left( {{2\,c_L} - {c_R}} \right)\,\sqrt {\frac{{{c_L}\,(2\,J - 3\,{c_L}\,{M^2})}}{{{c_R}\,(5\,{c_L} - 4\,{c_R})}}} \,.
\end{equation}
Imposing $\dot C>0$, one recovers the condition (\ref{eqPositiveTempCondition}) for local thermodynamic stability. In order to impose the Lloyd bound one has to consider also contributions from the rotation of the black hole, i.e. \cite{Cai:2016xho, Huang:2016fks}:
\begin{equation}
\dot C \le {(M - \Omega \,J)_ + } - {(M - \Omega \,J)_ - } \,,
\end{equation}
where the plus and minus signs indicate calculations on the outer and the inner horizons. The latter inequality leads to a lower bound on the mass of the black hole
\begin{equation}\label{eqLowerBoundontheMass}
M \ge \frac{{({2\,c_L}- {c_R})\,\sqrt {{c_L}} }}{{3\,\sqrt {12\,{c_R} - 15\,{c_L}} }}\,,
\end{equation}
which is true under condition (\ref{eqPositiveTempCondition}) for local thermodynamic stability. In terms of the original gravitational parameters this bound becomes
\begin{equation}
M \ge \frac{L}{3}\left( {1 - \frac{4}{{3 + {\nu ^2}}}} \right)\,.
\end{equation}
For $\nu\to 1$, the mass is zero, while for large $\nu \to \infty$, one has $M\geq L/3$.
\section{Conclusion}\label{sec:conclusion}\label{sec8}

Investigating the properties of various black hole solutions plays an important role in revealing hidden features of the elusive theory of quantum gravity. In this context, motivated by the remarkable duality between gravitational and gauge field theories, we study the thermodynamic properties of three-dimensional warped AdS black hole solution and its dual conformal field theory. Our findings uncover the candidates for proper Riemannian metrics on the space of equilibrium states of the WAdS$_3$ solution, together with several criteria for thermodynamic stability of the system. This allows us to constrain the possible values of the central charges from the dual field theory and the gravitational parameters from the bulk theory, thus making the duality more explicit. Our investigations were conducted mostly within the framework of thermodynamic information geometry, which utilizes the power of differential geometry to study statistical features of a given system. 
\\
\indent In general, the formalism of thermodynamic information geometry treats the space of equilibrium states as a Riemaninan manifold, equipped with a proper metric and an affine connection. Here, geodesic paths correspond to quasi static processes, while the distance between different macro states can be used to measure the probability for fluctuating between states. The parameter regions, where the geometric description breaks down, are the critical points of the system or its spinodal curves, where the corresponding thermodynamic information curvature becomes singular. In this case, from thermodynamic point of view, near a critical point the underlying inter-particle interactions become
strongly correlated and the equilibrium thermodynamic considerations are no longer applicable. There are several ways to construct thermodynamic information metrics. Below we summarize our results for the warped AdS$_3$ black hole.
\\
\indent In the Hessian formulation we analyzed the Ruppeiner and the Weinhold thermodynamic
metrics and showed that they are inadequate for the description of the WAdS$_3$ black hole equilibrium state space. This is due to the occurring mismatch between the singularities of the heat
capacity and the spinodal curves of the corresponding thermodynamic curvatures. Because both metrics are conformally related, with the temperature being the conformal factor, we have decided to construct another information metrics based on the Hessian of the Helmholtz and Gibbs free energies. The results also showed that Hessian thermodynamic metrics are not well-suited for our case. Furthermore, utilizing the properties of Gibbs and Helmholtz thermodynamic potentials, we have identified the regions of global TD stability in canonical and grand canonical ensembles, respectively, and showed that they both coincide with the local thermodynamic stability of the WAdS solution. The  critical temperature (\ref{eqSpinodalT}), which separates thermodynamic stability from the regions of TD instability of the system, was also calculated.
\\  
\indent On the other hand, in the case of Legendre invariant metrics, all considered thermodynamic metrics
successfully manage to incorporate the relevant Davies type critical points. Consequently they can
be considered as viable metrics on the equilibrium state space of the WAdS$_3$ black hole.
\\
\indent The third approach, based on conjugate thermodynamic potentials, also revealed complete agreement with the Davies critical points of the heat capacity. Here, the compact form of the thermodynamic curvatures, allowed us to fully determine the regions of flat, hyperbolic or elliptic information geometries on the space of the equilibrium states of the warped AdS$_3$ system.
\\
\indent All approaches were considered together with the conditions for local and global thermodynamic stability. The latter imposed several constraints on the central charges from the dual field theory, e.g. Eqs. (\ref{eqPositiveTempCondition}) and (\ref{eqLTDSCinSJSpace}). Several relations, namely Eqs. (\ref{eqSpinBound}), (\ref{eqBulkParamRelation1}) and (\ref{eqSOmegaLocalTDstability}), between the bulk gravitational parameters were also found. Additionally, in several cases, we were able to strengthen the local and global TD stability of the black hole by imposing Sylvester's criterion for positive definite metrics. 
\\
\indent One can go even further and calculate the logarithmic correction to the entropy of the black hole, due to small thermal fluctuations around its equilibrium configuration. In Ref. \cite{Das:2001ic}, it was found that for any thermodynamic system, satisfying the first law, one can write the corrected form of the entropy as
\begin{equation}
S' = S +\alpha \,\log (C\,{T^2}) +  \cdots \,.
\end{equation}
Here, $S$ and $C$ are the uncorrected entropy and heat capacity, while the constant $\alpha$ depends on the underlying micro states, and dots denotes higher order corrections. For example, if we choose to work in ($J, T$) space, the corrected heat capacity of the WAdS$_3$ black hole becomes\footnote{Note that for black holes, very close to extremality ($T\to 0$) or criticality ($T\to T_c$), the fluctuation analysis ceases to be valid due to large quantum fluctuations.} 
\begin{equation}
C'(J,\,T) = T\,\frac{{\partial S'(J,\,T)}}{{\partial T}} = C + \alpha \left( {2 + \frac{T}{C}\,\frac{{\partial C(J,\,T)}}{{\partial T}}} \right)\,.
\end{equation}
Local TD stability ($C'>0$) now requires $T<T_c$ and
\begin{equation}\label{eqLocalTDCorrected}
J>\frac{{27\,{{\left( {{c_L}\,\pi \,T - 1} \right)}^2}\,\left( {1 - {c_L}\,\pi \,T\,\left( {2 + ({c_R} - {c_L})\,\pi \,T} \right)} \right){\alpha ^2}}}{{2\,{c_L}\,c_R^2\,{\pi ^4}\,{T^2}}}\,,
\end{equation}
which reduces to $J>0$ at $\alpha=0$. On the other hand, global thermodynamic stability, imposed by the concavity $\partial^2 F'/\partial T^2<0$, leads to $T<T_c$, together with
\begin{equation}\label{eqGlobalTDCorrected}
J > \frac{{27\,\left( {1 - {c_L}\,\pi \,T\,\left( {2 + ({c_R} - {c_L})\,\pi \,T} \right)} \right){\alpha ^2}}}{{2\,{c_L}\,c_R^2\,{\pi ^4}\,{T^2}}}\,.
\end{equation}
The latter bound on the momentum $J$ is stronger than Eq. (\ref{eqLocalTDCorrected}). Hence, one concludes that the log-corrected thermodynamics of the WAdS solution respects the condition $T<T_c$ for global TD stability from the uncorrected case, while imposing an additional non-zero lower bound on the angular momentum $J$. 
\\
\indent Finally, we have calculated the complexity growth of the WAdS$_3$ black hole within the ''Complexity equals Action'' proposal. The natural requirement  that the complexity growth be a positive quantity turned out to be compatible with the condition (\ref{eqPositiveTempCondition}) for local TD stability. Meanwhile, imposing Lloyd's bound on the complexity growth, we have found a lower bound (\ref{eqLowerBoundontheMass}) on the mass of the black hole.

\section*{Acknowledgments}

The authors would like to thank Seyed Ali Hosseini Mansoori and Behrouz Mirza for the valuable comments on the manuscript. This work was partially supported by the following Bulgarian NSF grants \textnumero~DM18/1, \textnumero~DN18/1 and \textnumero~N28/5. T. Vetsov gratefully acknowledge the support by the Bulgarian national program ``Young Scientists and Postdoctoral Research
Fellows 2019''.

%\begin{appendix}

%\end{appendix}

%\begin{thebibliography}{99}

%\bibitem{TS:2017}

%\end{thebibliography}

%====================
%	Bibliography
%====================

\bibliographystyle{utphys}
\bibliography{TS-refs}

\end{document}